\makeatletter\let\ifGm@compatii\relax\makeatother
\documentclass[pre,aps,twocolumn,showpacs,floatfix,superscriptaddress]{revtex4}
\usepackage{graphicx}
\usepackage{amsmath, amsthm, amssymb}
\usepackage{epsfig}
\usepackage{latexsym}
\usepackage{bm}
\usepackage{color}

\def\beqr{\begin{eqnarray}}
\def\eqnr{\end{eqnarray}}
\def\beq{\begin{equation}}
\def\bc{\begin{center}}
\def\ec{\end{center}}
\def\eqn{\end{equation}}
\def\sgn{\mbox{sgn}}
\def\R{{\cal{R}}}
\def\alphap{\alpha^{\prime}}
\def\p{^{\prime}}
\def\epsilon{\varepsilon}
\def\rmp#1#2#3{{ Rev. Mod. Phys.} {\bf #1}, #2 (#3)}
\def\prl#1#2#3{{ Phys. Rev. Lett.} {\bf #1}, #2 (#3)}
\def\epl#1#2#3{{ Euro. Phys. Lett.} {\bf #1}, #2 (#3)}

\def\pre#1#2#3{Phys. Rev. E {\bf #1}, #2 (#3)}

\def\pra#1#2#3{Phys. Rev. A {\bf #1}, #2 (#3)}
\def\pnas#1#2#3{Proc. Natl. Acad. Sci. (USA) {\bf #1}, #2 (#3)}

\def\physa#1#2#3{Physica A {\bf #1}, #2 (#3)}

\def\natl#1#2#3{Nature (London) {\bf #1}, #2 (#3)}

\def\jasa#1#2#3{J. Acoust. Soc. Am. {\bf #1}, #2 (#3)}

\begin{document}

\title{A general mechanism  for the `$1/f$' noise}

\author{Avinash Chand Yadav}
\affiliation{School of Physical \& Mathematical Sciences, Central University of Haryana, Mahendergarh 123 031, India}

\author{Ramakrishna Ramaswamy}
\affiliation{School of Physical Sciences, Jawaharlal Nehru University, New Delhi 110 067, India}

\author{Deepak Dhar\footnote{Present address: Department of Physics, Indian Institute of Science, Education and Research, Dr. Homi Bhabha Road, Pashan, Pune 411 008, India}}

\affiliation{Department of Theoretical Physics, Tata Institute of Fundamental Research, Homi Bhabha Road, Mumbai 400 005, India}

\begin{abstract}

We consider the response of a memoryless nonlinear device that converts an input signal $\xi(t)$  into an output $\eta(t)$ that only depends on the value of the input at the same time, $t$. For input Gaussian noise with power spectrum $1/f^{\alpha}$, the nonlinearity modifies the spectral index of the output to give a spectrum that varies as $1/f^{\alphap}$ with $\alphap \neq \alpha$. We show that the value of $\alphap$ depends on the nonlinear transformation and can be tuned continuously.  This provides a general mechanism for the ubiquitous `$1/f$' noise found in nature.

\end{abstract}


\maketitle

In a very wide variety of natural systems, temporal fluctuations in observables are found to be long-ranged,  
with an approximately  $1/f^{\alpha}$ divergence in the power-spectrum at small frequencies $f$. When the exponent $\alpha$ lies between 1 and 2, this is generally termed `$1/f$'  noise, and understanding the origin of such fluctuations as well as  its ubiquity has been a long-standing problem in physics \cite{Dutta_1981}. First noted in 1925, in the spectra of current fluctuations in vacuum tubes \cite{Johnson_1925},   $1/f$ noise has been seen in systems as  different as river discharges \cite{Wang_2008, Thompson_2012}, loudness and pitch fluctuations in music \cite{Voss_1975}, solar flares \cite{Paczuski_2007}, volatility in stock market prices \cite{Stanley_1999},  DNA sequence \cite{Kaneko_1992, Voss_1992}, current fluctuations in biological ion channels and synthetic channels (nanopores) \cite{Siwy_2002}, hydration dynamics on the lipid membranes \cite{Yasuoka_2015},  neuronal  potentials and currents and electroencephalography (EEG) recordings \cite{Pettersen_2014}.

The difficulty of generating non-integer values of $\alpha$ for the noise exponent within linear
response theory has  contributed to keeping this problem somewhat enigmatic and unresolved. 
In the past nine decades that the problem has been extant, several different explanations have 
been proposed (see below), but none of these have been fully satisfactory.  The search for  a 
general mechanism of the $1/f$ noise was the main motivation behind the proposal of self-organized 
criticality (SOC) by Bak, Tang and Wiesenfeld in 1987 \cite{Bak_1987} that yields long-ranged 
correlations in time and thus can generate $1/f^{\alpha}$ dependence in the power spectrum of 
fluctuations.  Several SOC models have been studied from this viewpoint  \cite{Maslov_1999, ali, 
Dhar_2006, Davidsen_2002, Laurson_2005,Yadav_2012},  but since $1/f^{\alpha}$ noise may 
be found in models with only a small number of degrees of freedom, clearly self-organized 
criticality  is  not a {\it necessary} condition for obtaining $1/f$ noise. 

The $1/f^{\alpha}$ spectrum, with $\alpha \geq 1$ is not integrable near $f=0$, and the variance 
of the signal would be infinite, if there were no cutoffs.  In practice, one finds that if the signal is 
studied for a duration $T$, then if one doubles the duration, the observed mean of the signal 
appears to drift, and the variance of the signal  increases with $T$. Thus, if the lower cutoff 
on the frequencies is $\sim 1/T$, the net power in the signal increases as $T^{a}$, where $a = \alpha -1$. 

Our main observation in this paper is easily described.  We consider a discrete-time Gaussian 
stochastic process $\xi(t)$, which has the power-law spectral density  
\beq
S_{\xi}(f) = \dfrac{A}{f^\alpha},
\label{ps_in}
\eqn
with a lower-cutoff $1/T$.  The variance of the signal (which is also equal to the total power $P_{\xi}$ in the signal)  increases as $T^{\alpha -1}$ when the cutoff $T$ is increased.  We generate the output signal $\eta(t)$ by applying a instantaneous nonlinear transform to the input signal, 
\begin{equation}
\label{rofxi}
\eta(t) = {\mathcal R}[\xi(t)],
\end{equation}
where the function $R$ is typically sigmoidal. A representative example of such a transform is given by
\begin{equation}
{\mathcal R}(x) = \sgn(x) |x|^b , {\rm ~ with~~} 0 \leq b \leq 1.
\label{resp}
\end{equation}
Then the total power in the output process is  $P_{\eta}$ = $\langle |\eta(t)|^2 \rangle $. Clearly, $P_{\eta}  \sim [P_{\xi}]^b$, and thus  the output process $\eta(t)$ has  a different dependence of the total power on the cutoff frequency $1/T$ than the input process $\xi(t)$. 

From the above argument, namely that {\it the nonlinearity renormalizes the spectral index $\alpha$},
one may na\"ively conclude that the renormalized value of the spectral index is given by $\alphap 
- 1 = b(\alpha -1)$.  This is not so, since the coefficient of proportionality also changes when $T$ 
is changed. We argue below that
\begin{equation}
S_{\eta}(f)  \sim B \frac{1}{T^{\beta\p} f^{\alphap}}, {\rm ~for~ } 1/T < f \ll 1,
\end{equation}  
where $B$ is some constant, and we determine  the exponents $\beta\p$ and $\alphap$ as a function of input spectral exponent $\alpha$ and the device-dependent parameter  $b$.  The previous argument only implies the constraint
\begin{equation}
\alphap - \beta\p = 1 + b (\alpha -1). 
\label{expon1}
\end{equation}
This observation is rather elementary, but seems not to have noted earlier in this context, or with this level of generality. 

Mechanisms for $1/f$ spectra that have been suggested previously can be grouped into five classes, and these are briefly discussed below for comparison.
\begin{enumerate}

\item The explanation advanced most frequently involves treating the process as a superposition of many independent pulse-like events with power-law distribution for inter-pulse intervals or relaxation times \cite{Dutta_1981, Milotti_2002,  Amir_2011}.  This only transfers the problem to finding the reason for the power-law in the distribution of pulse sizes, or pulse intervals. 

\item In another common approach \cite{Montroll_1982}, the output signal is seen as a product of several independent or weakly dependent random variables. The resulting probability distribution is therefore log-normal, and this gives an inverse power law over a fairly wide frequency window.

\item Systems with a small number of  freedoms and with chaotic deterministic  evolution, with or without external noise \cite{Benmizrachi_1985, Kaulakys_2005} have non-trivial time-correlations. In specific cases, the origin of the 1/$f$ noise has been traced to the specific form of the evolution equations; this cannot be termed a general mechanism.

\item  A detailed description of the system using nonlinear partial differential equations with external boundary or bulk noise terms \cite{Grinstein_1992}  can be made that leads in  some cases to $1/f$. Again this is not general.

\item
It  has been observed \cite{heiwong1} that a nonlinear transform changes the exponent in numerical calculations of spectra, but this observation was not supported by analysis or a formula for the changed exponents \cite{heiwong2}. 
\end{enumerate}

\begin{figure}[t]
  \centering
  \scalebox{0.25}{\includegraphics{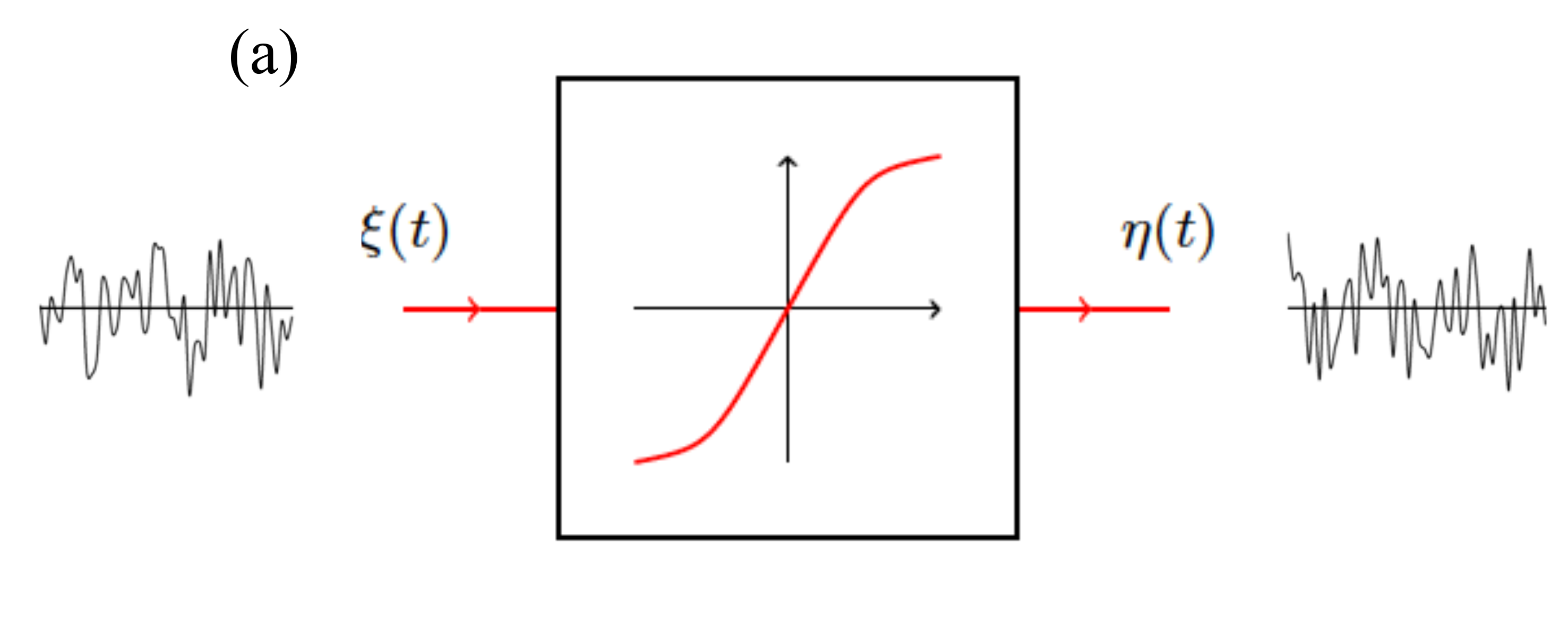}}
   \scalebox{0.54}{\includegraphics{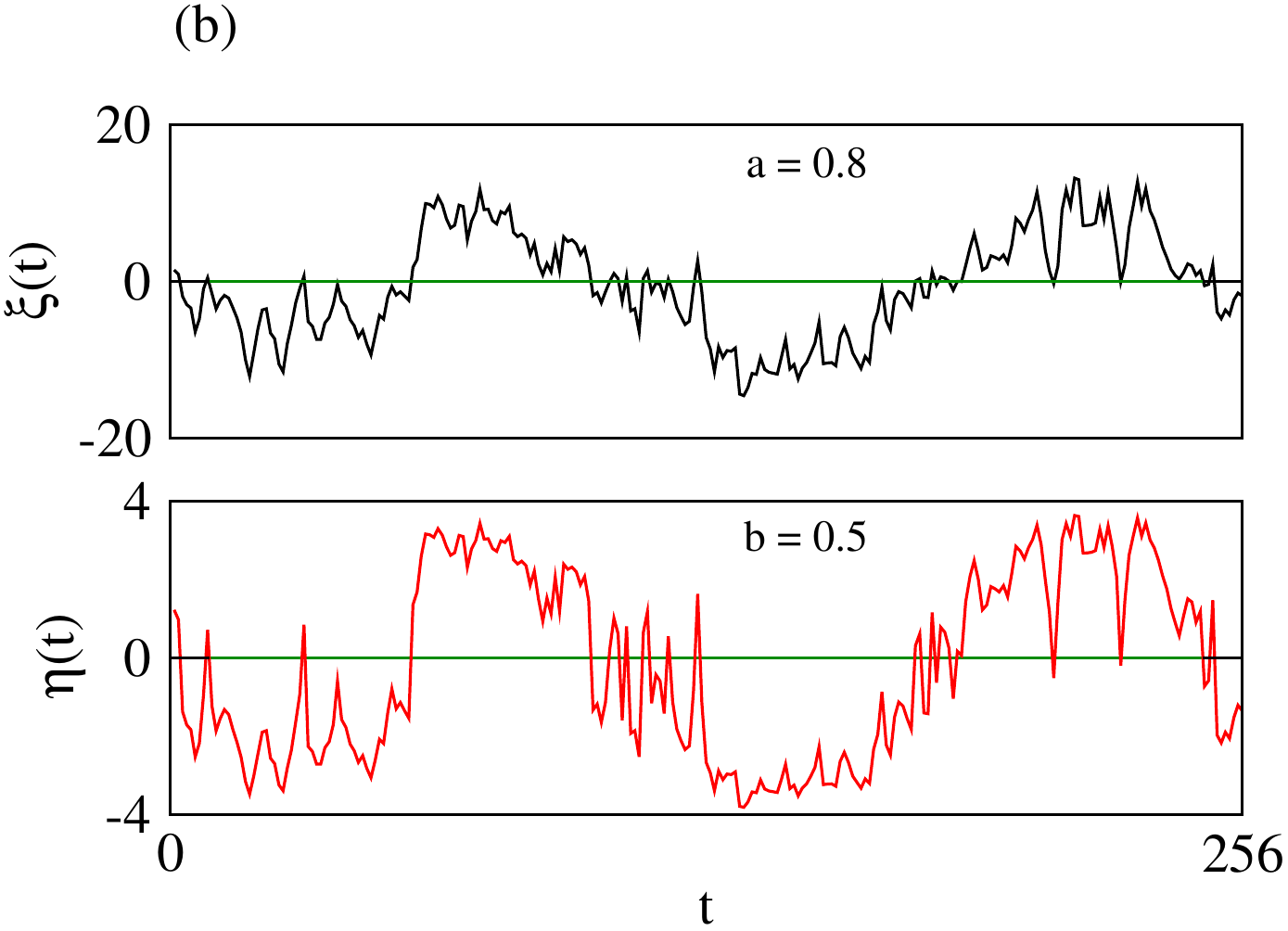}}
   \caption{(a) A noisy input signal $\xi(t)$ is fed into a device with nonlinear transfer function in order 
   to generate the output signal $\eta(t)$. (b) An example of the output signal $\eta(t)$ generated 
   from the input $\xi(t)$ with power spectrum $S_{\xi}(f) = 1/f^{1.8}$ and cutoff $T = 2^{8}$, 
   using the transformation $\eta(t) = \sgn[\xi(t)]|\xi(t)|^b$ for $b = 0.5$. }
\label{fig1}
\end{figure}

\begin{figure}[t]
  \centering
  \scalebox{0.65}{\includegraphics{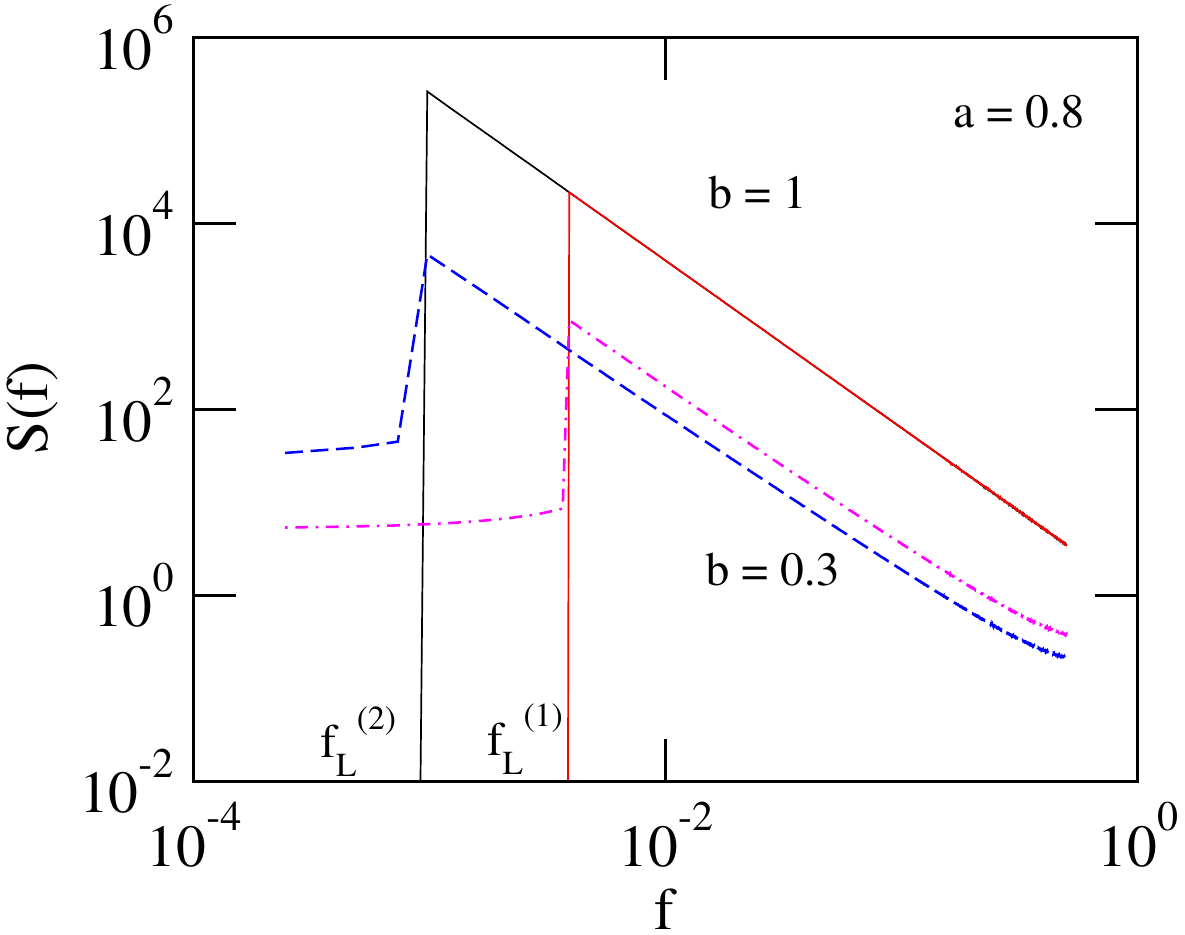}}
  \caption{Input (full lines) and output  power spectra (dotted lines) for two different values of the lower cutoff frequency.  The signal is of length $2^{12}$ and is averaged over $10^4$ samples. The lower cutoff 
  frequencies are $f_{L}^{(1)} = 2^4/2^{12}$ and $f_{L}^{(2)} = 2^2/2^{12}$. }
\label{fig2}
\end{figure}

In many examples of $1/f$ noise,  the response time of the system is much shorter than the time scale of observation.  Then, the nonlinear response of the system may be approximated as  a memoryless response, and this provides a general mechanism for generating the non-trivial power-law correlations seen in the time-series of different observables in these systems. In our earlier work  \cite{Yadav_2013}, we had noted this property when the input signal was bounded Brownian walk, but as we show here, this approach is significantly more general. 

Consider the input signal  $\xi(t)$ to be a discrete-time stationary Gaussian process, where $t$ takes integer values. From the discrete Fourier transform $\tilde{\xi}(f)$ of the signal in  a bounded  interval $[0,N]$ with $N\gg T$, namely
\begin{equation}
\tilde{\xi}(f) = \frac{1}{\sqrt{N}}\sum_{t=1}^{N}  \xi(t) \exp(2 \pi  i f t),
\end{equation}
one defines the power-spectrum $S_{\xi}(f)$ as
\begin{equation}
S_{\xi}(f) =  \langle |\tilde{\xi}(f)|^2 + |\tilde{\xi}(-f)|^2 \rangle.
\end{equation}
We will consider the case where $\tilde{\xi}(f)$ are independent complex Gaussian variables with zero mean and covariance $\langle \tilde{\xi}(f)\tilde{\xi}(f\p) \rangle = S_{\xi}(f)\delta_{ff\p}$.

Consider further that the input noise is fed into a {\em memoryless} nonlinear device, namely one where the output $\eta(t)$ is simply a nonlinear function of the instantaneous value of the input signal as represented in Eq.~(\ref{rofxi}).
For simplicity we take $\R(x)$ to be given by Eq.~(\ref{resp}) which is sigmoidal for $0 \le b \le1$ [see Fig.~\ref{fig1} (a)]. If the input signal is rescaled by multiplying by a factor $\lambda$, the output is rescaled by a factor $\lambda^b$; we use this rescaling to set $A$ = 1 in Eq.~(\ref{ps_in}). The transformation leaves the points of zero-crossing unchanged, but increases the steepness of the signal near the zero crossings, while the maxima and minima are made flatter; a typical transformed signal is shown in Fig.~\ref{fig1} (b).

Power spectra of input and output signals for two different lower cutoff frequencies are shown 
in Fig.~\ref{fig2}. The output power (at a fixed frequency $f$) decreases with decreasing cutoff 
frequency, although the total power in the output signal increases. Below the cutoff frequency,
the power spectrum of output is approximately constant, whose value increases when the 
cutoff frequency is decreased. The memoryless nonlinear transformation modifies the 
spectral index of power spectrum of the output as can be seen in Fig.~\ref{fig2}, and this can 
be explicitly computed as we show below. 

For the power spectrum of the input Gaussian signal considered here, the two point correlation function is of the form 
 \begin{equation}
\langle [\xi(t) - \xi(t+ \tau)]^2 \rangle \approx \begin{cases} 2 \mathcal{A}  |\tau |^{a}, & {\rm ~for~} 1  \ll \tau \ll T, \\  2 \mathcal{A} |T |^{a}, & {\rm ~for~}  \tau \gg T,\end{cases}
\end{equation}
where $\mathcal{A} = A/a$ for large $T$. The normalized two-point autocorrelation function of the process, $C(\tau)$, is 
\begin{equation}
C(\tau) =\frac{ \langle \xi(t) \xi(t+\tau) \rangle}{ \langle \xi(t)^2 \rangle}.
\label{eq1}
\end{equation}
For $\tau \gg T$, $C(\tau)$ tends to zero, and the expectation value $\langle \xi(t)^2 \rangle $ is  a finite value $  \mathcal{A} T^a$,  independent of $t$. With this normalization, $C(0)=1$. Writing $\epsilon = C(\tau)$, note
that for $\tau \ll T$, we have
\begin{equation}
\epsilon \approx 1 - \left| {\frac{\tau }{T}}\right|^a,
\end{equation}
and $\epsilon \to 0$ for $\tau \gg T$.

The covariance matrix
\begin{equation}
{\mathbb C} =  \left[ \begin{array}{cc} \langle \xi(t)^2 \rangle & \langle \xi(t) \xi(t+\tau) \rangle \\
\langle \xi(t) \xi(t+\tau) \rangle & \langle \xi(t +\tau)^2 \rangle   \end{array} \right],
\end{equation}
in terms of the parameters defined above can be written as 
\begin{equation}
{\mathbb C} =  \mathcal{A} T^a \begin{bmatrix} 1 & \epsilon \\
\epsilon & 1
\end{bmatrix}.
\end{equation}
Since $\xi(t)$ is a Gaussian process, the joint probability density that $\xi(t) \in [\xi_1, \xi_1 + d\xi_1]$ 
and $\xi(t+\tau)\in[\xi_2, \xi_2 + d\xi_2]$ is ${\rm Prob}(\xi_1,\xi_2) d\xi_1d\xi_2$, where  
the density ${\rm Prob}(\xi_1,\xi_2)$ is given in terms of the matrix ${\mathbb C}^{-1}$
and can easily be computed as 
\begin{equation}
{\rm Prob} (\xi_1,\xi_2) =  {\mathcal N} \exp  \left[- \frac{  \xi_1^2 + \xi_2^2 - 2 \epsilon \xi_1 \xi_2} {2 \mathcal{A} T^a ( 1 - \epsilon^2) } \right], 
\end{equation}
the normalization constant being
\begin{equation}
{\mathcal N} = 1/ [ 2 \pi \mathcal{A} T^a ( 1 - \epsilon^2)^{1/2}].
\end{equation}

The corresponding two-point correlation function for the output process $\eta(t)$,
$G(\tau) = \langle R(\xi(t)) R(\xi(t + \tau)) \rangle$, is 
\begin{equation}
G(\tau) = \int_{-\infty}^{+\infty}  d\xi_1  \int_{-\infty}^{+\infty}  d\xi_2 {\rm Prob}(\xi_1,\xi_2) R(\xi_1) R(\xi_2),\nonumber
\end{equation}
which can be reduced to the form
\beq
\label{eq15}
G(\tau) = 4 {\mathcal  N } \int_{0}^{\infty}  d\xi_1  \int_{0}^{\infty}  d\xi_2   [\xi_1  \xi_2]^b P(\xi_1,\xi_2),
\eqn
where
\beq
P(\xi_1,\xi_2) =   \exp  \left[- \frac{  \xi_1^2 + \xi_2^2 }  {2 A  T^a ( 1 - \epsilon^2) } \right] \sinh \left[ \frac{ \epsilon \xi_1 \xi_2}{ A T^a ( 1 -\epsilon^2)}\right]. \nonumber
\eqn

In order to simplify the analysis, we rescale the variables, $\bar{\xi}_j =\xi_j/[\mathcal{A}T^{a}(1-\epsilon^2)]^{1/2}$, $j=1, 2$;  Eq.~(\ref{eq15}) then reduces to 
\begin{equation}
G(\tau) = \frac{4  \mathcal{A}^b  T^{ab}  }{2 \pi} H_b(\epsilon),
\label{eqG12}
\end{equation}
where
\beq
H_b(\epsilon) = (1 -\epsilon^2)^{b+\frac{1}{2}} \int_{0}^{\infty}  d\bar{\xi}_1  \int_{0}^{\infty}  d\bar{\xi}_2~[\bar{\xi}_1 \bar{\xi}_2]^b Q(\bar{\xi}_1,\bar{\xi}_2),
\label{eq:h}
\eqn
and
\beq
Q(\bar{\xi}_1,\bar{\xi}_2) = \exp  \left[- \frac{  \bar{\xi}_1^2 + \bar{\xi}_2^2 }  
{2} \right] \sinh (\epsilon~\bar{\xi}_1\bar{\xi}_2). \nonumber
\eqn

Note that the $\tau$-dependence of the right hand side of Eq.~(\ref{eqG12}) comes entirely from the $\epsilon$-dependence of $H_b(\epsilon)$. For $\epsilon$ near 1 (i.e. small $\tau/T$) the singular dependence of $H_b$ on $\epsilon$ leads to a power-law tail in the power spectrum of $R[\xi(t)]$, and upon Taylor expansion we get
\begin{equation}
H_b(\epsilon) =     (1 - \epsilon^2)^{b+\frac{1}{2}}  \sum_{n=0}^{\infty}  \epsilon^{2n +1}  \frac{  [\Gamma(n +b/2+1)]^2 2^{2n+b} }{ (2 n + 1)!}.
\label{eqG}
\end{equation}

Using Stirling's approximation, the sum in the above equation can be shown \cite{stirling}
to have the form $\sum_{n=0}^{\infty} \epsilon^{2n} n^{b-1/2}$ which
 converges for $\epsilon^2  < 1$ while for $\epsilon^2$ near 1, it diverges 
as $(1-\epsilon^2)^{-b -1/2}$, exactly cancelling the multiplier. Therefore $G(\tau)$ tends 
to a finite limit as $\epsilon\to$ 1. This is to be expected since the $\epsilon \rightarrow$ 
1 limit is the same as $\tau \rightarrow 0$, and clearly $G(\tau =0)$ is finite; 
indeed
\begin{equation}
G(\tau =0) = \langle |\xi(t)|^{2b} \rangle = \mathcal{A}^b T^{ab} \frac{ 2^b \Gamma( b+1/2)}{\sqrt{ \pi}}. 
\end{equation}

The power-spectrum for $ 1/T \ll f \ll 1$ is determined by $G(\tau)$ for $ 1 \ll \tau \ll T$. In \cite{Yadav_2013} the case $\xi(t)$ considered was a bounded Brownian motion, with  $|\xi(t)| \leq  \sqrt{T}$. In that paper $\xi(t)$ is not Gaussian, but the difference process $\Delta \xi(t) = \xi(t+\delta) -\xi(t)$ is nearly Gaussian for $\delta \ll T$. For the transformation function $\R$ given by Eq.~(\ref{resp}), the autocorrelation function $C(\tau)$ of the output process is given by 
\begin{equation}
C(\tau) \sim 1 - K_1 \left|\dfrac{\tau}{T}\right| ^{b + \frac{1}{2}} - K_2 \left|\dfrac{\tau}{T}\right|  +{\rm ~higher ~order~ terms,}\nonumber
\end{equation}
for short times, $1 \ll\tau \ll T$, where $K_1$ and $K_2$ are constants and $ 0 \leq b \leq 1/2$. For the input process, the corresponding autocorrelation function is  $\langle \xi(t)\xi(t+\tau) \rangle_c \simeq 1 - |\tau/T| + {\rm ~higher ~order~ terms}$.  Assuming that the short-time correlations of the bounded Brownian motion are qualitatively the same as of a Gaussian signal with $\alpha$ = 2  and infrared cutoff, we make the identification
\begin{equation}
1 - \epsilon \simeq \left| \frac{\tau}{T}\right|,
\end{equation}
and conclude that in this case
\beqr
H_b(\epsilon) \approx  K_0 - K_1 (1- \epsilon)^{b+\frac{1}{2}} -K_2 (1-\epsilon) \nonumber\\ +~ 
{\rm higher~order~terms~ in~} (1 -\epsilon).
\eqnr

The  input noise has power-spectrum $1/f^{\alpha}$. Therefore $C(\tau) \approx 1 - |\tau/T|^{\alpha -1}$ giving
\begin{equation}
\epsilon = 1 - |\tau/T|^a,
\end{equation}
which leads to 
\beqr
G(\tau) \approx  G(0) -  G_1 \left|\frac{\tau}{T}\right|^{a\left(b+\frac{1}{2}\right)} - G_2 \left|\frac{\tau}{T}\right|^{a} + \dots,
\eqnr
$G_1$ and $G_2$ being constants.

Our final result therefore is that the leading order term in the autocorrelation function for the output process $\eta(t)$ varies as $|\tau/T|^{a(b+1/2)}$ for $ 1 \ll \tau \ll T$, and as $|\tau/T|$ for $a <$1, $b <$1/2.
\begin{figure}[t]
\centering
\scalebox{0.63}{\includegraphics{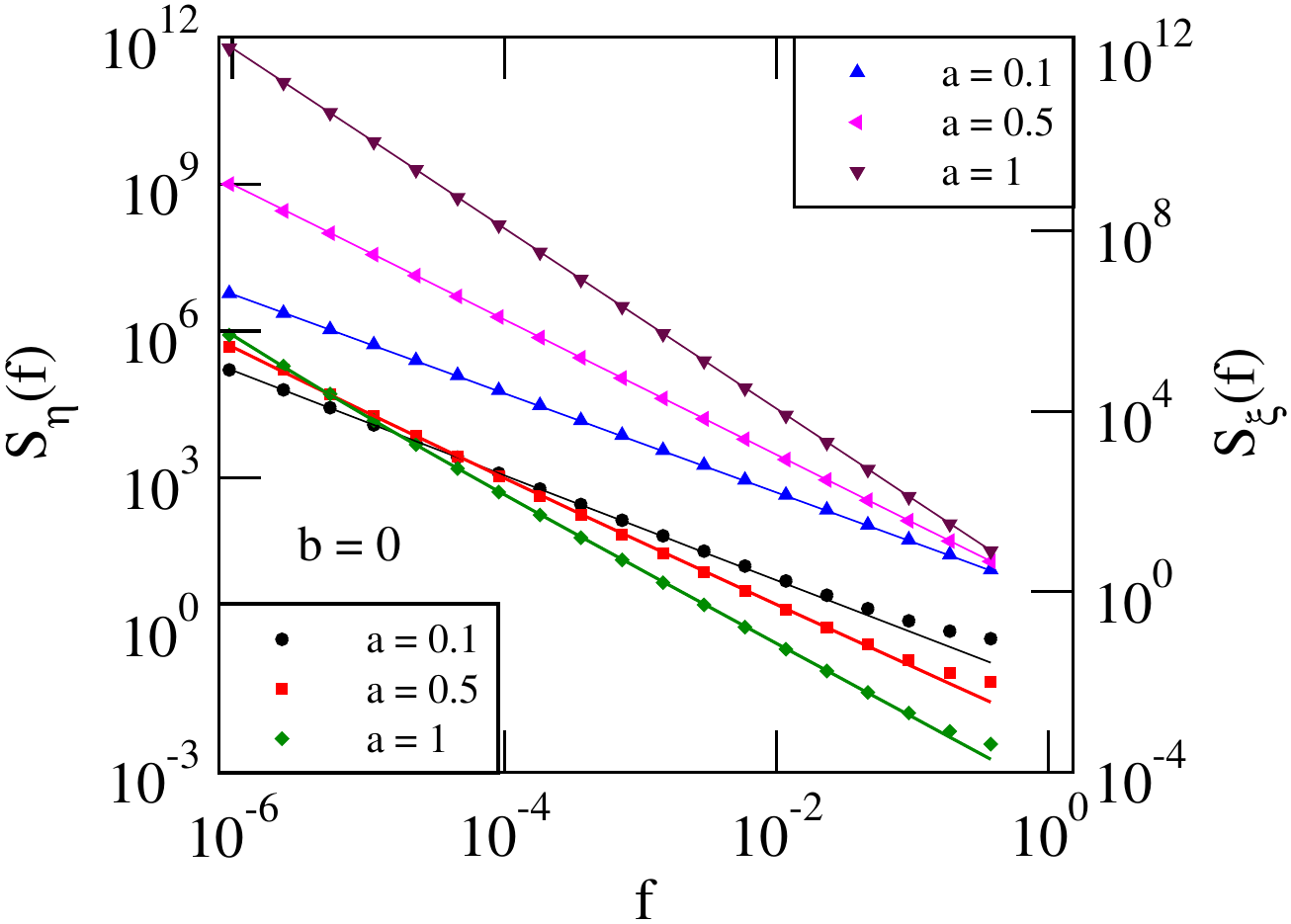}}
\caption{For different values of the parameter $a$, the power spectrum of the input signal $\xi(t)$ (top 3 plots) 
and the corresponding output signal $\eta(t)$ (bottom 3 plots) for $b = 0$, along with the best fit to the theoretical form.}
\label{fig3}
\end{figure}
As a consequence, when the input Gaussian process  $\xi(t)$ has a $1/f^{\alpha}$ power spectrum  
the output process $\R[ \xi(t)]$ has  a modified power spectrum $1/f^{\alphap}$. 
For the case of $\R$ given by Eq.~(\ref{resp}),
\beq
\alphap = \begin{cases} 1 + (\alpha -1)  (b+1/2), & {\rm for}~ 0 <b \leq 1/2, \\ \alpha, & {\rm for}~ 1/2 <b \leq 1,\end{cases}
\label{expon2}
\eqn
for $1 \leq \alpha \leq 2$, with $\alpha =  1+a$. From Eqs.~(\ref{expon1}) and 
(\ref{expon2}), we find that $\beta\p = (\alpha-1)/2$ for $0 <b \leq 1/2$.
 
These results can be verified by direct numerical studies. We generate $N= 2^{20}$ independent 
Gaussian random complex variables $\tilde{\xi}(f)$ with the variance of $\tilde{\xi}(f)$ varying as 
$1/f^{\alpha}$, where $f = j/N$ with $j = 0$ to $N-1$. The signal $\xi(t)$ is obtained by fast Fourier 
transformation (FFT), keeping only the real part. The series $\eta(t) = \R[\xi(t)]$ is then obtained, and 
its power spectrum is computed. (Further details are provided in the Supplementary Information).

The power spectra of the output signal for $b = 0$ with different $a$ are shown in Fig.~\ref{fig3}. 
A direct estimate of the exponent $\alphap$ from the slope of the log-log plot can be unreliable 
since corrections to scaling are significant. We therefore fit each spectrum to the theoretically 
expected form, $S(f) = A_1f^{-\alpha_1}+A_2f^{-\alpha_2}$, where $A_1$ and $A_1$ are constants, 
and $\alpha_1=1+a(b+1/2)$ and $\alpha_2 = 1+a$. In the frequency range $[10^{-5},10^{-1}]$, we
obtain excellent fits to the data, with $A_1 \gg A_2$, showing that the fitting form fits the data well.  
(More extensive tests are provided in the Supplementary Information).

To summarize, in the present work we have shown that memoryless nonlinear response of input Gaussian noise with power spectrum $1/f^{\alpha}$ produces noisy output that has power spectrum of the $1/f^{\alphap}$ type at low frequencies. The nontrivial value of the spectral exponent $\alphap$ depends both on the properties of the input noise, as well as on the characteristics of the transfer function.

ACY acknowledges the hospitality of the TIFR during several visits between 2013 and 2016. 
RR and DD are supported in part by the Department of Science and Technology (India) under the grants  
DST-SR/S2/JCB/2008  and DST-SR/S2/JCB-24/2005 respectively.


\begin{thebibliography}{99}

\bibitem{Dutta_1981} P. Dutta and P. M. Horn, \rmp{53}{497}{1981}; M. B. Weissman, \rmp{60}{537}{1988}.

\bibitem{Johnson_1925} J. B. Johnson, Phys. Rev. {\bf 26}, 71 (1925). 

\bibitem{Wang_2008} G. Wang, T. Jiang, R. Blender, and K. Fraedrich, J. Hydrol. {\bf 351}, 230 (2008).

\bibitem{Thompson_2012} S. E. Thompson and G. G. Katul, Adv. Wat. Res. {\bf 37}, 94 (2012).

\bibitem{Voss_1975} R. F. Voss and J. Clarke, \natl{258}{317}{1975}; \jasa{63}{258}{1978}.

\bibitem{Paczuski_2007} V. M. Uritsky, M. Paczuski, J. M. Davila, and S. I. Jones, \prl{99}{025001}{2007}; M. Baiesi, M. Paczuski, and A. L. Stella, \prl{96}{051103}{2006}.

\bibitem{Stanley_1999} Y. Liu, P. Gopikrishnan, P. Cizeau, M. Meyer, C. Peng, and H. E. Stanley, \pre{60}{1390}{1999}.

\bibitem{Kaneko_1992} W. Li and K. Kaneko, \epl{17}{655}{1992}.

\bibitem{Voss_1992} R. F. Voss, \prl{68}{3805}{1992}.

\bibitem{Siwy_2002} Z. Siwy and A. Fuli\'nski, \prl{89}{158101}{2000}. 

\bibitem{Yasuoka_2015} E. Yamamoto, T. Akimoto, M. Yasui, and K. Yasuoka, {Scientific Report} {\bf 5},  8876 (2015).

\bibitem{Pettersen_2014} K. H. Pettersen, H. Lind\'en, T. Tetzlaff, G. T. Einevoll, PLoS Comput. Biol. {\bf 10(11)}, e1003928 (2014).

\bibitem{Bak_1987} P. Bak, C. Tang, and K. Wiesenfeld, \prl{59}{381}{1987}.

\bibitem{ali} A. A. Ali, \pre{52}{R4595}{1995}.

\bibitem{Maslov_1999} S. Maslov, C. Tang, and Y. C. Zhang, \prl{83}{2449}{1999}; P. DeLosRios, M. Marsili, and M. Vendruscolo, \prl{80}{5746}{1998}.

\bibitem{Davidsen_2002} J. Davidsen and M. Paczuski, \pre{66}{050101(R)}{2002}.

\bibitem{Laurson_2005} L. Laurson, M. J. Alava, and S. Zapperi, J. Stat. Mech. L11001 (2005).

\bibitem{Dhar_2006} D. Dhar, \physa{369}{29}{2006}.

\bibitem{Yadav_2012} A. C. Yadav, R. Ramaswamy, and D. Dhar, \pre{85}{061114}{2012}.


\bibitem{Milotti_2002} E. Milotti, arXiv:physics/0204033v1.

\bibitem{Amir_2011} A. Amir, Y. Oreg, and Y. Imry, \pnas{109}{1850}{2011}.

\bibitem{Montroll_1982} E. W. Montroll and M. F. Schlesinger, \pnas{79}{3380}{1982}.


\bibitem{Benmizrachi_1985} T. Geisel, A. Zacherl, and G. Radons, \prl{59}{2503}{1987}; A. Ben-Mizrachi, I. Procaccia, N. Rosenberg, A. Schmidt, and H. G. Schuster, \pra{31}{1830}{1985}.

\bibitem{Kaulakys_2005} B. Kaulakys, V. Gontis, and M. Alaburda, \pre{71}{051105}{2005}.


\bibitem{Grinstein_1992} G. Grinstein, T. Hwa, and H. J. Jensen, \pra{45}{R559}{1992}. 

\bibitem{heiwong1} H. Wong, Microelectronics Reliability, {\bf 43}, 585 (2003).

\bibitem{heiwong2} H. Wong,  {\em A study of surface-related low-frequency noise in MOSFETs and metal films},  University of Hong Kong (1990),  available at http://tinyurl.com/j2c5l3t.


\bibitem{Yadav_2013} A. C. Yadav, R. Ramaswamy, and D. Dhar, \epl{103}{60004}{2013}.


\bibitem{stirling} 
For large $n$,  $[\Gamma(n +b/2+1)]^2 2^{2n}/ (2 n + 1)! \approx K n^{b-\frac{1}{2}}$
where $K$ is a constant.

\end{thebibliography}
\end{document}


\title{Supplementary Information for \\ 
A general mechanism for the $1/f$ noise \\ 
by Avinash Chand Yadav, Ramakrishna Ramaswamy, and Deepak Dhar}

\maketitle

Here we provide a detailed numerical results for the power spectra of the  output signals. 
The input signal $\xi(t)$ is constructed so as to have the power spectrum  
\begin{eqnarray}
\nonumber
S_{\xi}(f) &=& A f^{-\alpha}, {\rm ~for~} \frac{1}{T} \leq f \leq \frac{1}{2},\\
&=&  0 {\rm ~otherwise},
\label{input}
\end{eqnarray}
where $\alpha = 1 + a$ with $0\le a \le 1$ being a parameter characterizing input and the constant $A$ is set 1 throughout in our numerical calculations. To generate the output signal $\eta(t)$, a nonlinear 
transformation is employed,
 \begin{equation}
 \eta(t) = \mathcal{R}[\xi(t)],
 \end{equation}
with 
\begin{equation}
\mathcal{R}(x) = \sgn(x) |x|^b , {\rm ~ with~~} 0 \leq b \leq 1.
\end{equation}
Note that for $b=1$, the spectrum is unchanged. 

\begin{figure}[h]
  \centering
  \scalebox{0.60}{\includegraphics{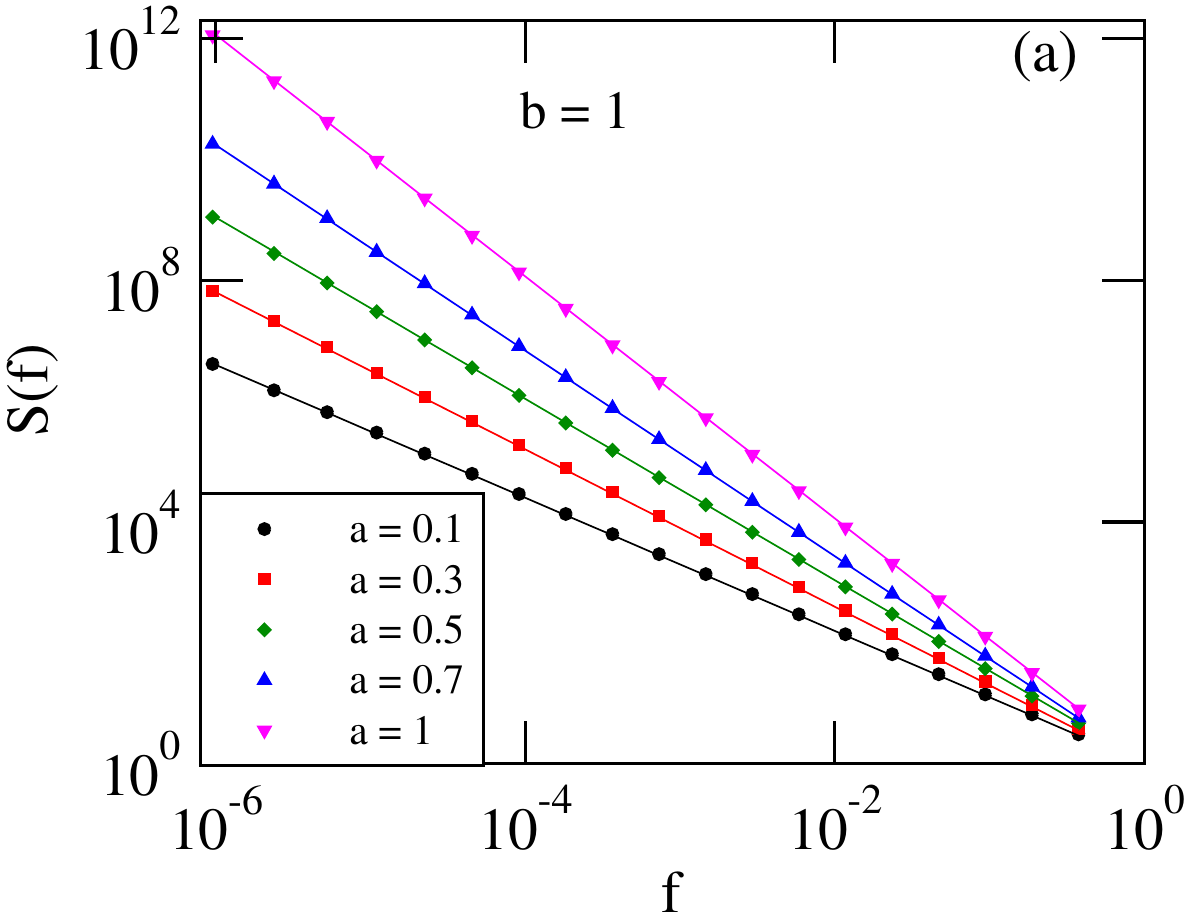}}
  \scalebox{0.60}{\includegraphics{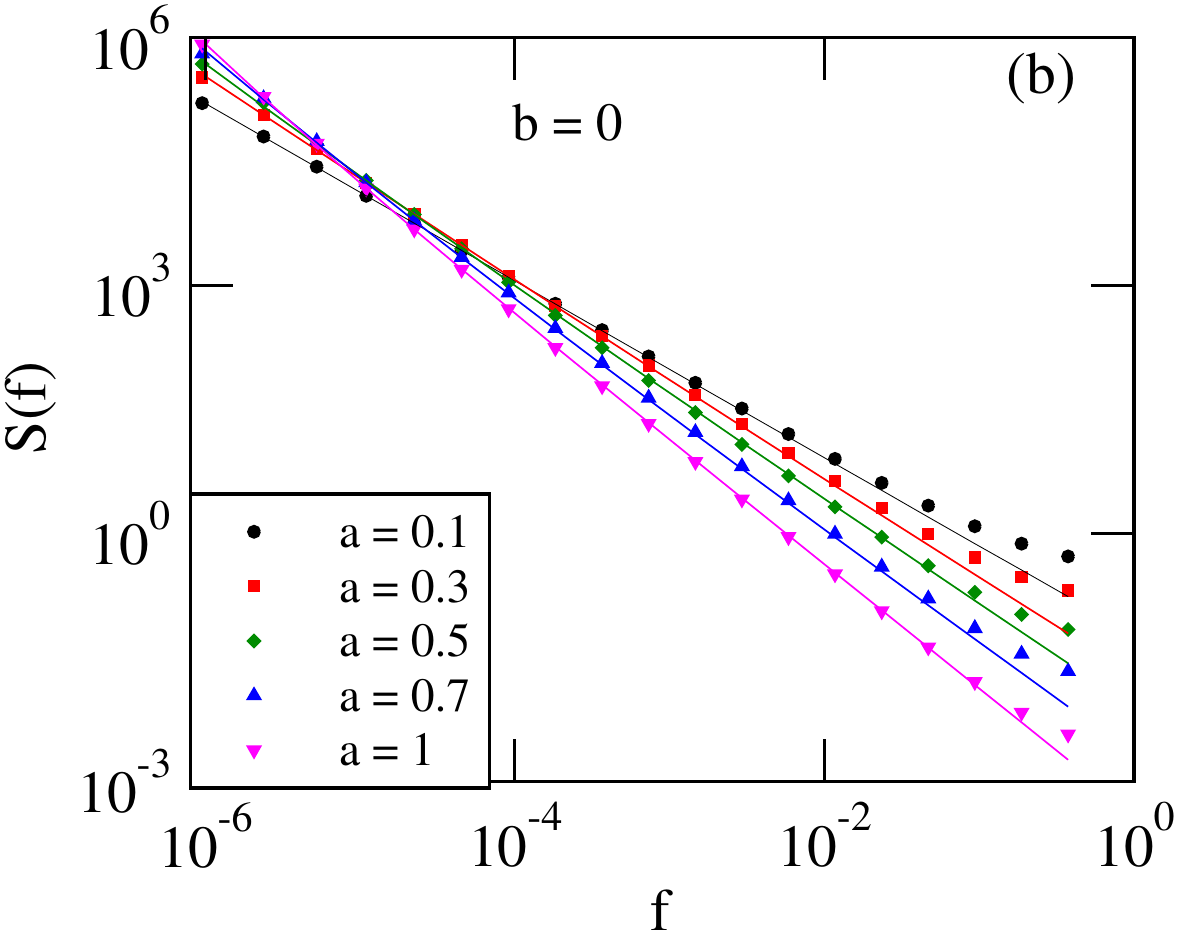}}
  \caption{Power spectra for the input ($b = 1$) and output ($b = 0$) at different parameter values $a$.  
  Continuous lines are the theoretically expected curves.  }
\label{ps0}
\end{figure}

The power spectra of the input ($b=1$) and output ($b=0$) signals for different values of $a$ are shown in 
Figure~S\ref{ps0}. The cutoff time is $T = 2^{20}$ and the power spectrum is ensemble averaged 
over $10^4$ independent realizations.  (Logarithmic binning is  used: 
the power spectrum is computed in frequency bins  $\Delta f= [2^{r}/N, (2^{r+1}-1)/N]$, where $N = 2^n$ 
and $r$ runs from 0 to $n-2$, and in the figures, the frequency is chosen as the average of lower and 
upper values of the frequency range in each bin.)

Each spectrum is fit to the theoretically expected form, 
\begin{equation}
S_{\eta}(f) = A_1f^{-\alpha_1}+A_2f^{-\alpha_2},
\end{equation}
where $A_1$ and $A_1$ are constants, and 
$\alpha_1=1+a(b+1/2)$ and $\alpha_2 = 1+a$. In the frequency range $[10^{-5},10^{-1}]$, we find good agreement with the theoretical form; see Tables S(\ref{tab1}) and (\ref{tab2}).

\begin{table}[h]
\centering
\begin{tabular}{ccccc}
\hline 

\hline
  $a$   & $\alpha_1$ & $\alpha_2$ & $A_1$ & $A_2$\\
\hline
 0.1 & 1.05 & 1.1 & 4.04 e-2 & 1.96 e-2 \\ 
 0.3 & 1.15 & 1.3 & 1.60 e-2 & 3.33 e-3 \\
 0.5 & 1.25 & 1.5 & 7.52 e-3 & 2.42 e-4 \\
 0.7 & 1.35 & 1.7 & 2.08 e-3 & 2.66 e-5 \\
 1.0 & 1.50 &  2.0 & 4.18 e-4 & 4.17 e-7 \\
\hline

\hline
\end{tabular}

 \caption{Numerically observed constants $A_1$ and $A_2$ for different values of $a$ with $b = 0$.  }
\label{tab1}
\end{table}

\begin{figure}[t]
  \centering
  \scalebox{0.60}{\includegraphics{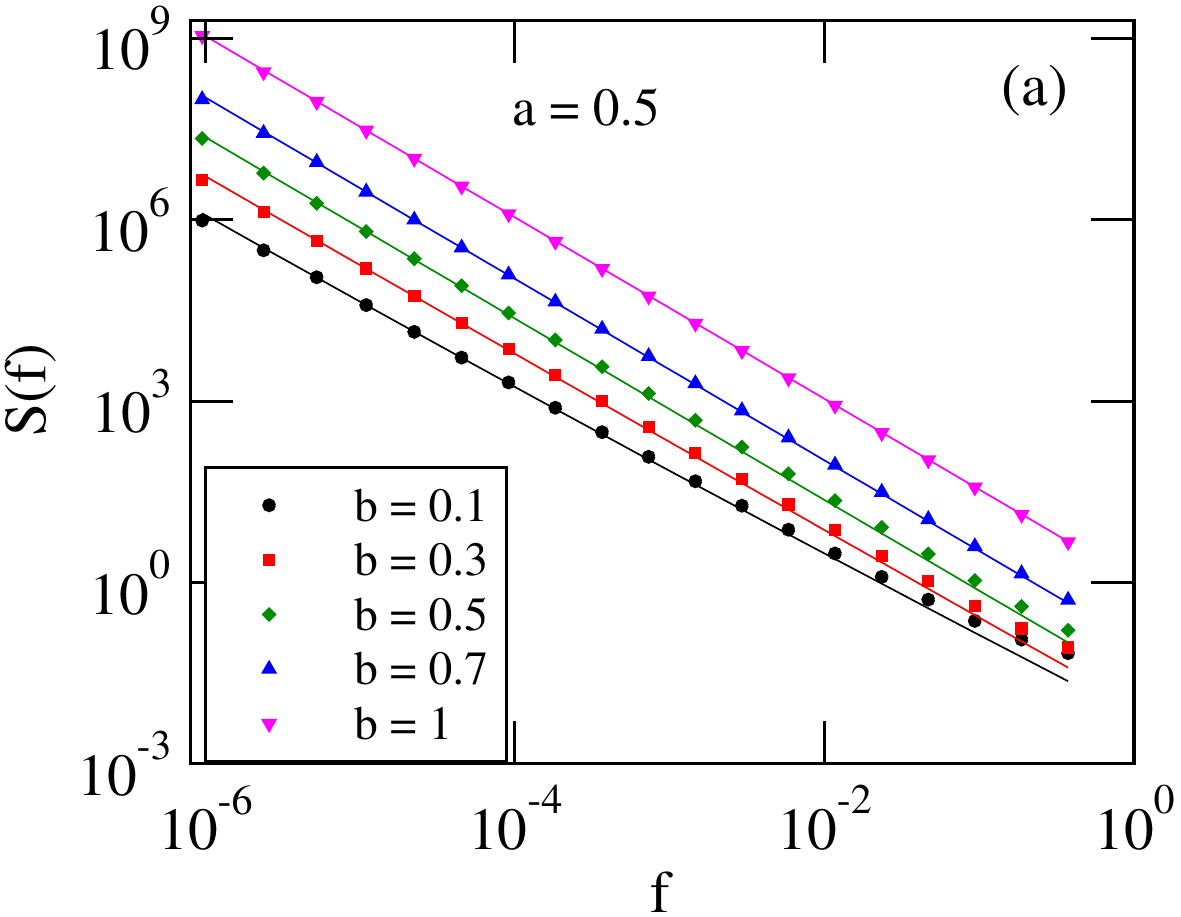}}
   \scalebox{0.60}{\includegraphics{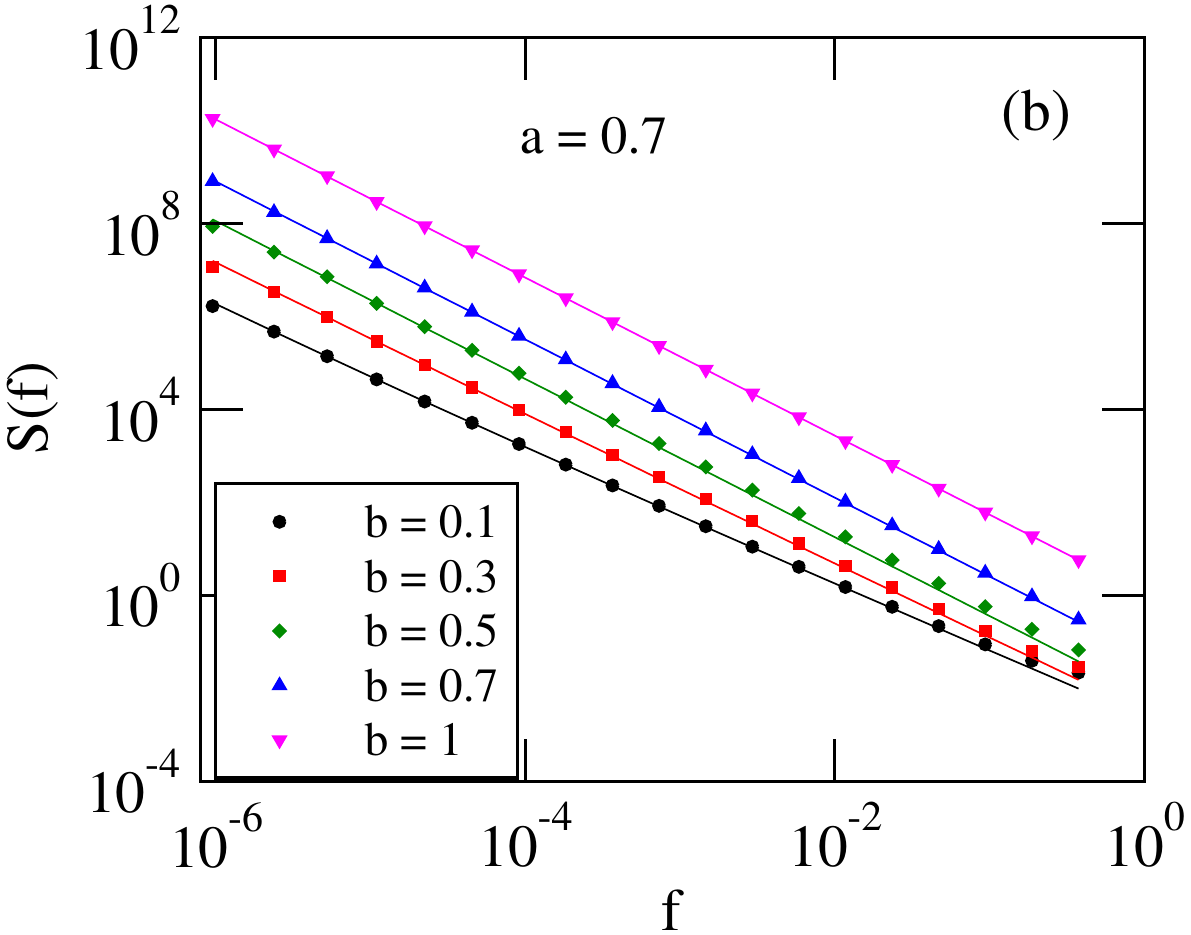}}
   \scalebox{0.60}{\includegraphics{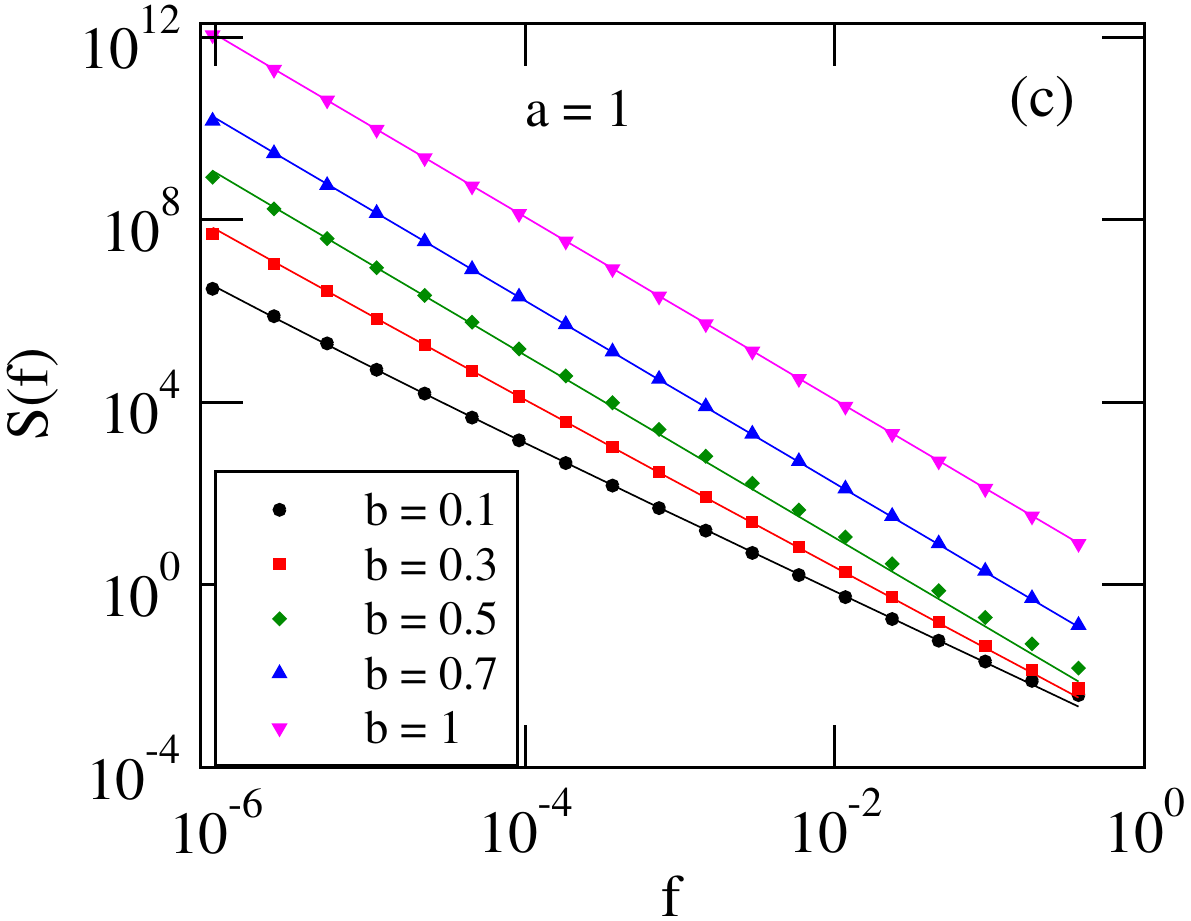}}
  \caption{Power spectra as a function of the parameter $b$ for different fixed values of  
  $a = 0.5, 0.7,$ and 1, along with the best fit theoretical form. }
\label{ps2}
\end{figure}

In Figure~S\ref{ps2}, we show similar plots for three different values of $a$: $ 0.5,0.7$ and 1, the 
last being used as an internal check of the numerical algorithms used in our computations. 

\begin{table}[t]
\centering
\begin{tabular}{ccccccc}
\hline 

\hline
  $a$ && $b$   & $\alpha_1 $ & $\alpha_2  $ & $A_1$ & $A_2$\\
\hline

\hline
0.5 && 0.1 & 1.3 & 1.5 & 5.45 e-3 & 8.52 e-4\\ 
0.5 && 0.3 & 1.4 & 1.5 & 5.67 e-3  & 3.81 e-3\\

0.7 && 0.1 & 1.42 & 1.7 &  2.35 e-3 & 6.76 e-5\\ 
0.7 && 0.3 & 1.56 & 1.7 &  2.68 e-3 & 5.25 e-4\\
    
 1 && 0.1 & 1.6 & 2 & 4.46 e-4 & 1.57 e-6\\ 
 1 && 0.3 & 1.8 & 2 & 5.40 e-4 & 2.75 e-5\\
 \hline

\hline
\end{tabular}
\caption{Numerically determined constants $A_1$ and $A_2$ for different 
values of $b$ with $a$ = 0.5, 0.7 and 1.  }
\label{tab2}
\end{table}

